# Guidance List for repOrting Bibliometric AnaLyses (GLOBAL): A Reporting Guideline Informed by a Two-Round Modified Delphi Study

Jeremy Y. Ng[1,2,3,4§] (0000-0003-0031-5873), Henry Liu[1] (0000-0001-9880-7248), Mehvish Masood[1] (0009-0008-2549-3573), Niveen Syed[1] (0009-0008-8375-3062), Dalia Zubashev[1] (0009-0008-5425-810X), Michel Sabé[5,6] (0000-0002-8530-5596), Marco Solmi[7,8,9,10,11] (0000-0003-4877-7233), Dimity Stephen[12] (0000-0002-7787-6081), Jens Peter Andersen[13] (0000-0003-2444-6210), Lutz Bornmann[14] (0000-0003-0810-7091), Rodrigo Costas[15] (0000-0002-7465-6462), Marianne Gauffriau[16] (0000-0001-7639-7719), Wolfgang Glänzel[17] (0000-0001-7529-5198), Sybille Hinze[18] (0000-0002-1729-5680), Cameron Neylon[19] (0000-0002-0068-716X), Nicolas Robinson-Garcia[20] (0000-0002-0585-7359), Alexander Schniedermann[12] (0000-0003-2132-7419), Gunnar Sivertsen[21] (0000-0003-1020-3189), Stephan Stahlschmidt[12] (0000-0003-3390-8632), Thed van Leeuwen[15] (0000-0001-7238-6289), Verena Weimer[22] (0000-0002-4500-7877), Simon Willemin[23] (0009-0003-5802-9113), Erija Yan[24] (0000-0002-0365-9340), Alesia Zuccala[25] (0000-0003-3045-0857), David Moher[1,10] (0000-0003-2434-4206), Ludo Waltman[15] (0000-0001-8249-1752), Stefanie Haustein[26,27,28] (0000-0003-0157-1430)

[1] Centre for Journalology, Ottawa Methods Centre, Ottawa Hospital Research Institute, Ottawa, Ontario, Canada
[2] Institute of General Practice and Interprofessional Care, University Hospital Tübingen, Tübingen, Germany
[3] Bosch Health Campus, Stuttgart, Germany
[4] Department of Health Research Methods, Evidence, and Impact, Faculty of Health Sciences, McMaster University, Hamilton, Canada
[5] Division of Adult Psychiatry, Department of Psychiatry, University Hospitals of Geneva, Thonex, Switzerland
[6] Faculty of Medicine, University of Geneva, Geneva, Switzerland
[7] Department of Psychiatry, University of Ottawa, Ontario, Canada
[8] Department of Mental Health, The Ottawa Hospital, Ontario, Canada
[9] Clinical Epidemiology Program, Ottawa Hospital Research Institute, University of Ottawa, Ottawa, Ontario, Canada
[10] School of Epidemiology and Public Health, Faculty of Medicine, University of Ottawa, Ottawa, Canada
[11] Department of Child and Adolescent Psychiatry, Charité Universitätsmedizin, Berlin, Germany
[12] German Centre for Higher Education Research and Science Studies (DZHW), Berlin, Germany
[13] Danish Centre for Studies in Research and Research Policy, Aarhus University, Aarhus, Denmark
[14] Administrative Headquarters of the Max Planck Society, Munich, Germany

[15] Centre for Science and Technology Studies, Leiden University, Leiden, The Netherlands
[16] IT University of Copenhagen, Copenhagen, Denmark
[17] KU Leuven, Leuven, Belgium
[18] Center for Open and Responsible Research, Berlin University Alliance, Berlin, Germany
[19] Curtin University, Perth, Australia
[20] University of Granada, Granada, Spain
[21] Nordic Institute for Studies in Innovation, Research and Education, Oslo, Norway
[22] DIPF | Leibniz Institute for Research and Information in Education, Frankfurt am Main, Germany
[23] ETH Zurich, Zurich, Switzerland
[24] Drexel University, Philadelphia, United States of America
[25] University of Copenhagen, Copenhagen, Denmark
[26] School of Information Studies, University of Ottawa, Ottawa, Canada
[27] Scholarly Communications Lab, Ottawa/Vancouver, Canada
[28] Centre Interuniversitaire de Recherche sur la Science et la Technologie, Université du Québec à Montréal, Montreal, Canada

[§]Corresponding Author Details:

Name: Dr. Jeremy Y. Ng, MSc, PhD
Address: The Ottawa Hospital, General Campus, Centre for Practice Changing Research Building, 501 Smyth Road, PO BOX 201B, Ottawa, ON, K1H 8L6 Canada
Email: ngjy2@mcmaster.ca

Abstract Word Count: 199
Manuscript Word Count: 5774

## Abstract

**Background**: Despite the growth of bibliometric analyses, few studies offer guidance on how to report them, resulting in a lack of research transparency and completeness. To address this gap, we developed the **Guidance List for the repOrting of Bibliometric AnaLyses (GLOBAL)**.

**Methods**: An initial list of items for the GLOBAL was generated through a scoping review and refined through a modified two-round Delphi, as recommended by the EQUATOR Network. The guideline was finalized through a pilot test, where participants assessed their manuscripts using the checklist and Elaboration and Explanation (E&E) document, which outlines item expectations and provides examples.

**Results:** The scoping review resulted in 32 items. In Round 1, 24 items reached consensus and thematic content analysis resulted in one additional item. This item, along with the eight items that did not reach consensus, were discussed in detail in Round 2. During the in-person meeting, one item was split into two, totalling ten voting items. Eight out of ten items reached consensus, five for inclusion and three for exclusion.

**Conclusions:** The finalized 28-item GLOBAL (and its accompanying E&E document) provides users with a framework to report bibliometric analyses, aimed at promoting high-quality reporting of bibliometric analyses in scholarly literature.

## Introduction

Bibliometrics is an established research field with its own journals and conferences. Bibliometrics is also a methodology applied in studies from a wide range of research fields from natural sciences and medicine to social sciences and humanities. Both in the bibliometric research field and in other research fields, the number of bibliometric analyses is growing. The **Guidance List for the repOrting of Bibliometric AnaLyses (GLOBAL)** presented in this paper is an evidence-informed reporting guideline aimed at promoting high-quality reporting of bibliometric analyses in scholarly literature.

The term bibliometrics is not defined unambiguously and several terms exist, for example, scientometrics and informetrics. The coverage of one of the terms often overlaps with that of others. We follow a pragmatic definition of bibliometrics based on common usage in the literature. We define bibliometric analyses as any study that quantitatively studies academic research based on at least one of two basic elements: (1) publications (e.g., journal articles, conference proceedings papers, books or book chapters, preprints, peer review reports, or grey literature) to represent scholarly outputs; and (2) citations (i.e., formal references to a publication in the reference lists of other publications) to reflect connections and knowledge flows between publications. In the context of research evaluation, citations are often used to indicate impact of individual articles, authors, journals, institutions or countries. Units of measurement of these elements can be applied to various entities and at various levels of aggregation, for instance: micro-level (e.g., authors, documents), meso-level (e.g., institutions, departments, journals), and macro-level (e.g., countries, disciplines). Some examples of these units are provided in **Table 1**.

Bibliometric analyses may introduce, adapt, and/or apply various types of bibliometric indicators, ranging from absolute numbers of publications and citation rates (e.g., journal impact factor, field-normalized citation rate) to citation percentiles or collaboration strength. These indicators are used to measure, compare and benchmark entities (AlRyalat et al., 2019; Donthu et al., 2021; Sugimoto & Larivière, 2018). Researchers and organizations conduct bibliometric analyses for a variety of purposes, such as to explore the intellectual structure of  fields, and to identify publication-related characteristics, trends, and patterns specific to journals, articles, books, authors, institutions, and/or topics of study. The findings of bibliometric analyses can serve to advance the knowledge about research activities by providing an overview of the research conducted, understanding how research has evolved over time, identifying knowledge gaps, and inspiring novel ideas for investigation in that particular area (Donthu et al., 2021).

Recently, a number of articles have been published describing how to report a bibliometric analysis (Ng et al., 2024a). However, most of these articles have not framed their work in the format of a reporting guideline (Jappe, 2020; Donthu et al., 2021; Linnenluecke et al., 2020) or do not follow the rigorous methodologies outlined by EQUATOR (EQUATOR Network, n.d.a; Moher et al., 2010) for developing reporting guidelines, such as conducting a literature review and a face-to-face consensus meeting. Such articles ranged from opinion pieces (Cabezas-Clavijo et al., 2021) to those with partial or complete literature reviews but lacking a consensus process (Boyack et al., 2022; Cabezas-Clavijo et al., 2023; Koo & Lin., 2023). Moreover, while one article, the Guideline for Reporting Bibliometric Reviews of the Biomedical Literature (BIBLIO) (Montazeri et al., 2023), fully adhered to EQUATOR's toolkit, gaps in reporting—such as limited details on the literature review process and missing descriptive statistics from Delphi voting rounds—make it difficult to assess its validity and reliability. The current lack of evidence-based guidance on how to report a bibliometric analysis can be problematic for several reasons. If authors fail to provide readers with enough information about how and when their study was conducted, including the database from which the bibliographic data was retrieved, readers will only have a partial understanding of what was done. Consequently, insufficient reporting may hinder the reproducibility of a study and further inhibit researchers from evaluating the accuracy of its findings or using the findings in new research (Bornmann et al., 2021). Furthermore, editors and peer reviewers have no guideline against which to compare the reporting quality of a study under their consideration. Moral and ethical justifications also exist for providing accurate research reporting (Moher, 2007). Ethical research promotes knowledge, truth, and the avoidance of error, which are values that are essential to both collaborative work and accountability to the public (Resnik, 2015). This holds especially when results of bibliometric analyses play a role in research evaluation and might therefore affect institutional funding allocations or individual careers. This article, therefore, describes the development of the GLOBAL, the first guideline developed through international multi-stakeholder consensus with the aim to enhance the quality of reporting bibliometric analyses in scholarly literature.

The GLOBAL is formatted to support the reporting process of research articles intended to be published in scholarly journals or on preprint servers. The design of the GLOBAL incorporates the conventional sections of "abstract", "introduction",

"methods", "results", and "discussion", along with an "other" section. We aimed to ensure that this reporting guideline is clear and easy to follow, as recommended by the Consolidated Standards of Reporting Trials (CONSORT) (Altman, 1996) "[r]eaders should not have to infer what was probably done; they should be told explicitly." Although the GLOBAL aims to enhance the reproducibility of bibliometric analyses, we acknowledge that practical considerations (e.g., journal requirements or concision) may prevent researchers from providing the full scope of information needed to meet these ideal standards of reporting for transparency.

# Methods

## Open Science Statement

The GLOBAL is registered on the EQUATOR Network Library of Reporting Guidelines (EQUATOR Network, n.d.b). The protocol was registered on January 12, 2023, on the Open Science Framework (OSF) at https://osf.io/nvu6w (Ng et al., 2023). Anonymized, aggregate voting data and participant responses from rounds 1 and 2 were also shared publicly at https://osf.io/mtxbf/files/k97h8. Identifying information was not shared. Participants in both Delphi survey rounds provided consent to participate in this study. We followed the Accurate Consensus Reporting Document (ACCORD) checklist (Gattrell et al., 2024) in reporting our study.

## Study Design

The following section briefly outlines the study design, while detailed explanations are provided in the subsequent sections. **Figure 1** provides a summary of the methods section.

The GLOBAL was developed in accordance with the EQUATOR Network's methodological framework (EQUATOR Network, n.d.a; Moher et al., 2010). First, a scoping review was conducted to identify any existing reporting guidance for bibliometric analyses and generate a preliminary list of candidate items for the GLOBAL. This scoping review has been posted as a preprint (Ng et al., 2024b) and published in a journal (Ng et al., 2025). The preliminary list of the GLOBAL candidate items was further developed using a two-round modified Delphi conducted on a global scale. In general, the Delphi method is used to formally obtain information and achieve consensus among experts regarding a particular topic. It is an iterative method in which multiple rounds of discussion build on each other until consensus is reached, based on a pre-defined level of formal agreement among participants (Murphy et al., 1998). This Delphi is considered "modified" because the preliminary list of items was generated through a scoping review and consultations with the GLOBAL steering committee, rather than originating directly from the Delphi panel, as is the standard practice. However, participants had the opportunity to suggest new items during both rounds. The GLOBAL steering committee consisted of six international researchers, four with expertise in bibliometrics (SHaustein, LW, MSabé, and MSolmi) and two with expertise in reporting guidelines (JYN and DM) to ensure the development of a robust guideline. Round 1 of the Delphi involved individuals completing an online survey using Welphi (*Welphi*, n.d.), a web-based platform that is specifically designed to host surveys employing the Delphi method. Round 2 consisted of an in-person consensus group meeting with participants who completed the previous round and were interested and able to attend the meeting. The penultimate guideline that emerged from this two-step process was then pilot tested for clarity and finalized.

## Scoping Review and Candidate Item Generation

An initial list of candidate items for the GLOBAL was generated through a scoping review (Peters et al., 2020) of peer-reviewed literature, articles on preprint servers, and grey literature that aimed to identify and categorize bibliometric reporting recommendations. Full details of the scoping review are available in Ng et al. (2025). Twenty-three studies met the inclusion criteria following screening. Consensus on the inclusion of an item, the section the item belonged to (i.e., “title”, “abstract”, “introduction”, “methods”, “results”, “discussion”, or “other” section of the reporting guideline), and the phrasing of candidate items for the GLOBAL were decided after multiple discussions with the steering committee and research team (HL, MM, and NS). The steering committee also had the opportunity to add items which they believed were necessary to increase the quality of bibliometric reporting but were not yet addressed by the included studies (Ng et al., 2025). This process resulted in a 32-item preliminary checklist; 31 items were created based on recommendations from the literature and one item arising from expert opinion of the GLOBAL steering committee (Item 3.8: Describe the data cleaning methods, including any limitations).

## Recruitment of Delphi Participants

Participants were recruited from a diverse group of international stakeholders with bibliometric expertise (e.g., bibliometricians, librarians, journal editors, policy and research analysts, and researchers) through purposeful sampling to ensure representation and generalizability. Respondents were required to speak and understand English to participate in this study. Steering committee members did not participate in either Delphi rounds but were responsible for panellist invitation and selection. Recruitment was conducted through two methods. First, the steering committee compiled a list of experts from the bibliometric community and sent personalized email invitations and reminders (see https://osf.io/mtxbf/files/ab465) to these potential participants through the Welphi platform (*Welphi*, n.d.). Second, an advertisement and recruitment script with a general universal link to the Welphi survey was disseminated to members of the International Society for Scientometrics and Informetrics (ISSI) through their mailing list and an ISSI website blog post on 26 July 2024 (Ng, Liu, & Haustein, 2024), and promoted via social media (X, LinkedIn). Information on the GLOBAL Delphi was also listed on the website for the 2024 International Science, Technology and Innovation Indicators (STI) conference [see https://sti2024.org/sti-conference/global-delphi-survey/] (STI, 2024). The survey link for both methods (i.e., the personalized recruitment email and the universal link) led participants to a page that provided more information about the study, including data privacy/storage information (see https://osf.io/mtxbf/files/pvesh). By completing the survey, participants provided consent to take part in the study. Participants were not provided financial compensation for completing the study. Those who participated in Round 2 of the Delphi were invited to co-author the present paper.

## Delphi Round 1

In Round 1, participants completed an online Delphi survey (see https://osf.io/mtxbf/files/pvesh for the survey) on the Welphi platform (*Welphi*, n.d.). The survey was open from 10 July 2024 to 16 August 2024, with reminder emails sent to participants who received personalized email invitations one, two, and four weeks following the initial email. Prior to administration, the survey was pilot tested from 29 June 2024 to 4 July 2024 by four researchers (DS and three external research assistants mentioned in the acknowledgements). Pilot testers did not participate in the Delphi. The pilot test was conducted to check for issues in survey design, technology, and clarity/phrasing of the survey questions.

The survey included 41 questions that addressed the following: (1) demographic variables (Ontario Agency for Health Protection and Promotion [Public Health Ontario], 2021) (seven close-ended questions); (2) preferences for GLOBAL candidate items (32 questions); and (3) other comments (two open-ended questions; suggestions for new items that were not addressed in the GLOBAL and additional comments in general). All survey questions were optional to complete, with the exception of rating preferences for the candidate items. Participants were required to complete all the questions on a page to move to the next, but their responses were submitted even if the survey was not fully completed. For the 'preferences for GLOBAL candidate items' section, participants were asked to rate each item of the preliminary 32-item GLOBAL guideline that was generated from the scoping review (Ng et al., 2025) using the following Likert scale scoring system (Jebb et al., 2021): essential (1-3), preferable (4-6), and non-essential (7-9). It was determined that items that received 80% of responses in the top range (7-9) or bottom range (1-3) on the 9-point scale were considered to have achieved consensus for inclusion or exclusion, respectively. This 80% threshold was selected based on general agreement within the literature, which commonly uses 75% as a threshold to define consensus (Diamond et al., 2014), and to solidify our findings. Items that met consensus in Round 1 for either inclusion or exclusion were considered final and were not rated again in Round 2. Each candidate GLOBAL item in this section also had an open-ended comment box for respondents to provide further feedback. At the end of the survey, participants were provided with information regarding the Round 2 in-person consensus meeting and a linked form to express their interest in attending.

## Delphi Round 2

A one-day consensus meeting was held on 21 September 2024 in Berlin, Germany, to discuss and vote on items that did not reach consensus or were newly introduced through Round 1 (see https://osf.io/mtxbf/files/pvesh for the survey questions). The date and location of the meeting were selected to coincide with the Science, Technology and Innovation Indicators (STI) 2024 conference held in Berlin from September 18 to 20, 2024, thereby facilitating the participation of many members of the bibliometric community. Stakeholders were invited by the steering committee via email from the list of Round 1 Delphi participants who fully completed the survey and expressed interest in participating in Round 2. A total of 32 participants were invited to participate. Efforts were made to ensure varied representation from stakeholder groups.

The in-person consensus meeting was moderated by three steering committee members (JYN, SHaustein, and LW), who did not vote or participate in discussions and maintained neutrality throughout the meeting. Two researchers (DS and one external research assistant mentioned in the acknowledgements) took notes and recorded votes during this process. In addition to taking notes, the meeting was recorded and transcribed using MacWhisper (*MacWhisper*, n.d.). During the meeting, all items from the initial literature review that did not reach consensus and all new items proposed by participants in Round 1 were discussed. The consensus group participants were presented with each item and its score from the first Delphi round, along with any remarks made by participants on that item, both six days in advance of the meeting (15 September 2024) and during the meeting itself on 21 September 2024. Participants at the consensus meeting were asked to comment on the significance of each item and whether it should be included in the GLOBAL. Following open discussion of an item, participants were given the opportunity to suggest revisions to its wording, which were implemented only if supported by a majority of the group. Subsequently, an anonymous electronic vote on the rephrased item was held using Mentimeter (*Interactive Presentation Software - Mentimeter*, n.d.), with the option to 'include in checklist', 'exclude from checklist', and 'abstain from voting' for each given item. After voting on an item, the results were presented quantitatively. Similar to Round 1, the inclusion and exclusion threshold of 80% served to represent majority consensus (Diamond et al., 2014). Items that did not reach consensus in either direction were excluded. Participants also had the chance to suggest new items for the GLOBAL during the consensus meeting which were subsequently voted on in the same session. While participants were encouraged to stay for the complete Delphi meeting, they were able to leave early, particularly due to international travel.

## Delphi Results Analysis

Frequencies and percentages were used to record the number of participants that completed each round of the Delphi and their basic demographic characteristics. For Round 1, qualitative data (open-ended responses) underwent thematic content analysis (Marks & Yardley, 2004). There were three categories of open-ended responses from the Round 1 survey: 1) item-specific responses (32 questions); 2) suggestions for new items that were not addressed in the GLOBAL (one question);

and 3) additional comments in general (one question). Coding to identify common themes in participant responses was conducted independently and in duplicate by two researchers (MM and NS). Following this, MM and NS met to iteratively generate and discuss themes and subthemes until consensus was reached. All 'item-specific responses' were reviewed and discussed, but only items deemed to have sufficient data, as determined by team discussion (JYN, HL, MM, and NS), were analysed (e.g., items that had less than three dissimilar comments were determined to have insufficient data). Item-specific responses were coded with the purpose of identifying ways to rephrase items on the GLOBAL for the Round 2 consensus meeting and to capture any concerns regarding GLOBAL items. Responses for 'suggestions for new items that were not addressed in the GLOBAL' were coded with the intention to identify new items to add to the GLOBAL (see **Table 2** notes for newly added items). Newly proposed items were subsequently presented to the research team and steering committee for further refinement. Through iterative team discussions, new items reached agreement to include in Round 2. Responses from 'additional comments in general' were used to generate general themes regarding participant preferences on the GLOBAL's format and usage and were subsequently presented to participants taking part in the Round 2 Delphi.

## Guidance List Pilot

Following the consensus meeting, a pilot of the penultimate checklist was conducted to evaluate the clarity, comprehensiveness, and real-world applicability of the GLOBAL. Researchers who were in the process of writing, submitting, or peer-reviewing bibliometric analyses were invited to participate in the pilot-test. The pilot-test was promoted through  blog posts on ScholCommLab on 7 August 2025 (see

https://www.scholcommlab.ca/research/global/) and Leiden Madtrics on 12 August 2025 (see https://www.leidenmadtrics.nl/articles/making-bibliometric-reporting-more-transparent-and-consistent-participants-needed-for-pilot-testing-of-global). Additionally, JYN sent invitation emails to Delphi participants and those who emailed their interest in completing the pilot test on 13 August 2025, while SHaustein reached out to other professional networks on the same day. The recruitment scripts/emails can be found at https://osf.io/mtxbf/files/6tfpj.

Participants received the checklist, the accompanying Explanation & Elaboration (E&E) document (Ng et al., 2026), which includes examples and explanations that clarify and justify each checklist item, and a Google feedback survey form, which was open from 6 August 2025 to 1 October 2025. Pilot-testers were asked to evaluate their manuscripts using the checklist and E&E document, and rate the clarity of each reporting item and the helpfulness of the related E&E text. The survey also allowed participants to suggest improvements to the items and/or explanations through an open-ended question (see https://osf.io/mtxbf/files/9rp32 for the survey questions).

Quantitative feedback was analysed using frequencies and percentages, while open-ended suggestions underwent thematic content analysis, with codes and themes summarized for the steering committee to review, by two research assistants (NS and DZ). The steering committee then reached consensus on final modifications to item wording and explanations to enhance clarity and usability. The finalized checklist and E&E document aim to improve the transparency and completeness of bibliometric reporting in scholarly literature.

# Results

## Deviations from the Protocol

There were three deviations from the protocol. First, the three-round Delphi process was reduced to two rounds as all the items except two reached consensus within the two rounds. Second, although Round 2 was initially planned to be an online survey, it was conducted as an in-person consensus meeting instead. While this deviation limited the number of participants that were able to attend Round 2, it promoted a productive and detailed discussion on the necessity of each item. Third, Welphi (*Welphi*, n.d.) was used to create and distribute the Round 1 Delphi survey instead of SurveyLet (*Calibrum*, n.d) because Welphi was designed to implement the Delphi method and ensure data accuracy.

## Round 1

### Participants

A total of 145 participants, representing 111 institutions (see https://osf.io/mtxbf/files/k97h8 for the complete list of institutions), took part in Round 1 by rating at least one GLOBAL item. Only two (1.4%) participants did not complete the survey in full. Most respondents were men (n = 91; 62.8%) and between the ages of 35 and 44 (n = 56; 41.4%). More than half of participants identified as ‘White’ (n = 80; 59.7%), while others selected ‘East Asian’ (n = 17; 12.7%), ‘Latino’ (n = 9; 6.7%), ‘South Asian’ (n = 6; 4.8%), or ‘Middle Eastern ‘(n = 5; 3.7%). Respondents worked in various countries, including the United States (n = 19; 16.0%), Canada (n = 15; 12.6%), Germany (n = 9; 7.6%), the United Kingdom (n = 9; 7.6%), and the Netherlands (n = 8; 6.7%). The top five roles reported by participants who completed the survey were ‘bibliometrician’ (n = 48; 22.6%), ‘librarian/information specialist’ (n = 36; 17.0%), professor (n=35; 16.5%) [including ‘associate professor’ (n = 18; 8.5%) and ‘full professor’ (n = 17; 8.0%)], and ‘research coordinator’ (n = 13; 6.1%). More than half of respondents reported to have more than ten years of experience in their respective roles (n = 70; 57.9%), a quarter had five to ten years of experience (n = 31; 25.6%), and 14.1% had less than five years (n = 17). **Table 3** provides a summary of participant demographics. Anonymized, aggregated participant responses and voting data from rounds 1 and 2 are available at https://osf.io/mtxbf/files/k97h8 and https://osf.io/mtxbf/files/jmx7g, respectively.

**GLOBAL Items**

A total of 24 out of 32 items reached the 80% consensus threshold for inclusion in the GLOBAL following Round 1. Content analysis of participant feedback resulted in one novel candidate item for inclusion in the GLOBAL: ‘Provide a clear study materials and data sharing statement (e.g., if datasets, data sources, codes used for the analysis, software, and/or calculations are provided or not)’. This item was voted on in Round 2. Within the open-ended survey responses, participants suggested two themes: 1) expanding the GLOBAL objective by adapting it to different audiences and/or types of records that use bibliometric analyses; and 2) reformatting the GLOBAL to reduce redundancy, include examples, or rephrase existing items (see **Table 4** for a summary of the themes and https://osf.io/mtxbf/files/7a8j9 for the complete coding and thematic content analysis). The first theme relates to the GLOBAL extensions discussed and encouraged in the ‘Future Directions’ section, while the second theme could not be implemented, as participants were only allowed to vote on the necessity of the items in reporting bibliometric analyses rather than modifying their content. A summary of voting results is provided in **Table 2**.

## Round 2

### Participants

A total of 16 participants took part in Round 2, with two (n = 12.5%) leaving early. Demographic characteristics were collected from Round 1, anonymized, and aggregated. Participants were mostly men (n = 11; 68.8%), 'White' (n = 13; 81.3%), and between 30 and 59 years of age (n = 16; 100%). Most respondents worked in Germany (n = 4; 25.0%), Denmark (n = 3; 18.8%), and the Netherlands (n = 2; 12.5%). Common roles included 'bibliometrician' (n = 7; 26.9%), 'journal editor' (n = 3; 10.7%), 'librarian/information specialist' (n = 3; 10.7%), and 'associate professor' (n = 3; 10.7%). Within these roles, participants had more than ten years (n = 10; 62.5%), between five and ten (n = 3; 18.8%), and less than five years (n = 1; 6.4%) of work experience. Summarized participant demographics are provided in **Table 3.**

### GLOBAL Items

Participants voted on ten items during the in-person consensus meeting. Initially, there were nine items from Round 1 that required further discussion, eight of which did not reach consensus and one that was introduced after content analysis of participant feedback. However, one item was split into two during Round 2 (see notes on modifications made in **Table 2**), resulting in ten items. The phrasing of seven items was altered. Five out of ten items reached the 80% consensus threshold for inclusion following Round 2, with all five undergoing rephrasing. The other five items were excluded, with three of them reaching consensus for exclusion. The remaining two items did not reach the consensus threshold and were therefore excluded. Participants did not suggest any new items for the GLOBAL. **Table 2** provides a summary of voting results, including notes on all the changes that occurred during each round. In total, 29 out of 34 items reached consensus for inclusion in the GLOBAL following the completion of both Delphi rounds. **Figure 2** provides a summary of the results section.

## Post-Delphi Item Modifications

Following the Delphi and during the finalization of the reporting guideline, discussions with steering committee members and Round 2 Delphi participants resulted in a few changes to the GLOBAL items. Due to overlapping reporting expectations regarding data sharing and accessibility, the items "Describe the availability and accessibility of data" and "Provide a statement on whether study materials, data, and/or code are shared, and if so, where and how they can be accessed" were merged into a single item: "Describe the availability and accessibility of data by providing a statement about whether study materials, data, and/or code are shared, and if so, where and how it can be accessed." Additionally, the word "citations" was removed from the item "Use references and citations to support statements and methods used," as citations can refer to both outgoing (sources cited in your own work) and incoming (instances when others cite your work) citations in bibliometrics, potentially causing confusion. In contrast, "references" specifically refers to citations in your own work to other sources, making it a clearer and more appropriate term for this item. Finally, the order of the items under the “others” section of the checklist was adjusted to improve the flow of reporting, aligning with the structure commonly found in research papers. A summary of these changes is provided in **Table 5**.

## Guideline Finalization Pilot

A total of 12 experts participated in the pilot test, with half (n = 6.50%) acting as authors of a manuscript reporting a bibliometric analysis and the remaining (n = 6.50%) serving as reviewers of a manuscript reporting a bibliometric analysis. Most participants indicated that the items were clear and that the corresponding E&E sections were helpful to understanding the item; a more detailed summary is available at https://osf.io/mtxbf/files/dbp8z. Within the open-ended survey responses, participants suggested two themes: (1) reword the GLOBAL items and E&E document to enhance clarity and applicability, and (2) reformat the GLOBAL to reduce redundancy and streamline items (see **Table 6** for a summary of the themes). While some feedback was incorporated into the E&E document, most suggestions could not be implemented, as there were not enough responses supporting the change or were specific to modifying or removing checklist items, which was not possible after the iterative Delphi process that resulted in these items. Anonymized data and complete coding and thematic content analysis can be found at https://osf.io/mtxbf/files/c4snk.

The finalized GLOBAL is comprised of 28 items, with the following in each section: “abstract” (one item), “introduction” (four items), “methods” (thirteen items), “results” (four items), “discussion” (three items), and “other” (three items). The finalized GLOBAL is presented in **Table 7** and is now available to download (see https://osf.io/mtxbf/files/5acmn) for any researcher to use for the reporting of bibliometric analyses.

# Discussion

The GLOBAL serves as the first guideline developed for the reporting of bibliometric analyses in scholarly literature through international multi-stakeholder and multi-sector consensus. Through a community-driven and iterative, multi-step process, we have developed a 28-item reporting guideline that is intended to enable more thorough and transparent reporting of bibliometric analyses. It is important to note that these are minimum standards. Authors should not feel discouraged from including additional information that might enhance the quality of reporting of their bibliometric analysis.

## Scope of the GLOBAL

The goal of the GLOBAL is to provide minimal but essential guidance for the reporting of bibliometric analyses in scholarly literature. The intent of the GLOBAL is not to provide methodological design guidance for researchers and specialists conducting bibliometric analyses, nor does it assess the suitability of particular methods in specific contexts. However, while our proposed work does not directly address the quality of bibliometric analyses, we anticipate that this reporting guideline will set the stage for future work in this area. The complete reporting of novel or more specialised types of bibliometric analyses may require additional guideline items and authors should not be deterred from reporting this information, but the GLOBAL should nevertheless be considered as base guideline by such studies, until necessary specialised extensions are developed. The latter may also address the reporting of other 'metrics' associated with bibliometrics (e.g., the reporting of webometrics or altmetrics).

## Implications of the GLOBAL

With the widespread use of bibliometric methods, establishing reporting guidelines is crucial to strengthening the reliability and accuracy of bibliometric studies. We expect such guidelines to enhance the quality of the reporting of research by assisting authors to ensure published papers are complete and transparent, thereby positively influencing how researchers plan, execute, and report their work (Donthu et al., 2021; Gagnier et al., 2013; Moher et al., 2010).

The GLOBAL has the potential to benefit many stakeholders. Thorough reporting, supported by adherence to reporting guidelines, allows readers to evaluate the usefulness of a study's methods and, consequently, the reliability and robustness of its conclusions. High-quality reporting may help to ensure new research is efficiently used and less research waste is produced, and may also facilitate easier replication and potential review updates (Moher et al., 2010). As a reporting guideline, the GLOBAL aims to assist researchers and peer reviewers in reporting and evaluating bibliometric analyses published in the scholarly literature.

## Implementation and Dissemination

The GLOBAL E&E document is designed to facilitate the use of the GLOBAL. It includes clear examples from published bibliometric literature that satisfy the items, providing authors with a framework on how to approach the reporting of their analyses. Additionally, it includes explanations that outline the essential components of each item, along with the rationale behind their inclusion. The E&E document of the GLOBAL was modelled after similar documents created for other reporting guidelines, such as CONSORT (Altman, 1996). It incorporates feedback from Delphi participants and steering committee members and draws on relevant peer-reviewed literature. The document underwent multiple rounds of revision to improve its clarity and usability. It can be found at the following link https://doi.org/10.21203/rs.3.rs-9322986/v1.

We plan on disseminating the guidelines via multiple sources, including but not limited to the following: 1) the core bibliometrics and reporting guidelines communities via conferences and/or mailing lists associated with the International Society for Scientometrics and Informetrics (ISSI) (https://www.issi-society.org/), the International Network of Research Management Societies (INORMS) (https://inorms.net/), the Directorate for Science, Technology and Innovation (STI) (https://www.oecd.org/sti/), the STI conference series (https://enid-europe.eu/index.php?id=confSeries), the European Network of Indicator Designers (ENID) (https://enid-europe.eu/), and the International Congress on Peer Review and Scientific Publication (https://peerreviewcongress.org/); 2) editors and editorial board members of scholarly journals; 3) researchers from disciplines that use bibliometrics and/or reporting guidelines to evaluate their own fields; 4) scholarly communication librarians and research managers that conduct bibliometric analyses to support researchers; 5) publishers and publishing-related organizations/associations that publish bibliometric analyses or bibliometric-related studies; 6) websites and blogs that feature bibliometric-related content, such as The Scholarly Kitchen (The Scholarly Kitchen, n.d.) and Leiden Madtrics (Leiden Madtrics, n.d.); 7) developers of applications and software that assist researchers unfamiliar with bibliometric analyses in conducting and reporting them, such as Bibliometrix (Bibliometrix, n.d.); and 8) schools offering bibliometric-related programs, such as the European

Summer School for Scientomerics (ESSS) (European summer school for scientometrics, n.d.), the Centre for Science and Technology Studies (CWTS) Scientometrics Summer School (CWTS, n.d.), the German Centre for Higher Education Research and Science Studies (DZHW, 2025), Robert K. Merton Center for Science Studies (RMZ) (Robert K. Merton Center for Science Studies, n.d.), and the Scholarly Communications Lab (ScholCommLab) (ScholCommLab, n.d.).

Future studies may include GLOBAL extensions that address the reporting of other “metrics” associated with bibliometrics, such as webometrics and altmetrics. Additionally, while the current focus is on reporting in the scholarly literature, such as in journal articles, it would be valuable to develop reporting guidelines for other types of bibliometric analyses, such as analyses performed for research institutions, research funders, governments, and other stakeholders, for instance in a research assessment context. Future research could also explore the development of reporting guidelines for studies that use bibliometrics along with other methodological approaches, such as systematic reviews.

Further research may also examine the facilitators and barriers to the use of the GLOBAL by authors, editors, and peer reviewers, and develop interventions to overcome identified barriers and evaluate those interventions. Moreover, conducting think-aloud studies, where participants verbalize their thoughts while completing a task, can help assess how items are interpreted. Reliability studies to identify where items can be interpreted differently would also be beneficial to inform potential revisions to the guideline (Charters, 2003).

Translations of the reporting guideline will improve the accessibility of the GLOBAL. We encourage journal editors and publishers to promote the GLOBAL (for instance, by mentioning it in their journals' "Instructions to Authors" page), endorse its usage, advise editors and peer reviewers to assess submitted bibliometric analyses against the GLOBAL, and make adjustments to journal policies to take into account the new reporting recommendations.

## Strengths and Limitations

This study has several strengths. First, the development of the GLOBAL adheres to recommendations present within the EQUATOR toolkit and other established guidelines for developing a reporting guideline (EQUATOR Network, n.d.a.; Moher et al., 2010), thereby increasing its robustness. Second, the development process is evidence-based, supported by a comprehensive scoping review (Ng et al., 2025) of recommendations in the literature. Third, involving diverse stakeholders from the international community (e.g., researchers with varying years of experience with bibliometrics) in the selection process strengthens the study's credibility and relevance as it considers a wide range of perspectives. Fourth, the recruitment of participants through two methods, sending personalized emails and issuing an open invitation through public advertisement, helped to minimize the potential for bias that could arise from selecting individual participants or relying on a single sampling method.

Conversely, weaknesses of the study include a possible decrease in representativeness due to English-language restrictions, which limited participation by non-English speakers (Khanna et al., 2022). The in-person character of the consensus meeting in Berlin is another limitation, as not all stakeholders were able to attend the meeting and provide feedback on the GLOBAL items, thereby restricting participant diversity.

## Conclusions

The GLOBAL serves as a guide for high-quality reporting of bibliometric analysis in scholarly literature. We anticipate that the GLOBAL will be useful to bibliometricians, librarians, policy and research analysts, and researchers, as well as authors, editors, and peer reviewers of bibliometric analyses. Ultimately, the goal of the GLOBAL is to promote more thorough and transparent reporting of bibliometric analyses.

## List of Abbreviations

EQUATOR: Enhancing the QUAlity and Transparency Of health Research

GLOBAL: Guidance List for the repOrting of Bibliometric AnaLyses

ISSI: International Society for Scientometrics and Informetrics

OSF: Open Science Framework

## Declarations

## Ethics Approval

Research ethics approval was obtained by the Ottawa Health Science Network Research Ethics Board (REB ID #20230527-01H).

## Consent for Publication

All authors consent to this manuscript's publication.

## Availability of Data and Materials

All relevant data are included in this manuscript (Delphi participant demographics, Delphi themes, Delphi participant voting results, and checklist) or posted on the Open Science Framework (recruitment script email, Delphi item questions, Delphi participant demographics, Delphi participant voting results, and checklist): https://doi.org/10.17605/OSF.IO/MTXBF.

## Competing Interests

MSolmi received honoraria/has been a consultant for Angelini, AbbVie, Boehringer Ingelheim, Lundbeck, Otsuka, all unrelated to this work. All other authors declare that they have no competing interests.

## Funding

JYN's postdoctoral fellowship was funded by a MITACS Elevate Award (Award #: IT36020), and co-funded by EBSCO Health. We also gratefully acknowledge funding provided by Cabells, Clarivate, and the Korean Institute of Oriental Medicine. The funders played no role in the study design and conceptualization, data collection and analysis, decision to publish, or preparation of this research.

## Authors' Contributions

JYN: designed and conceptualized the study, collected and analysed data, co-drafted the manuscript, and gave final approval of the version to be published.

HL: collected and analysed data, co-drafted the manuscript

MM: collected and analysed data, co-drafted the manuscript

NS: collected and analysed data, co-drafted the manuscript

DZ: collected and analysed data, co-drafted the manuscript

MSabé: provided input for the design and methodology, made critical revisions to the manuscript, and gave final approval of the version to be published.

MSolmi: provided input for the design and methodology, made critical revisions to the manuscript, and gave final approval of the version to be published.

DS: provided input for the design and methodology, assisted with the collection and analysis of data, made critical revisions to the manuscript, and gave final approval of the version to be published.

JPA: assisted with the collection and analysis of data, made critical revisions to the manuscript, and gave final approval of the version to be published.

LB: assisted with the collection and analysis of data, made critical revisions to the manuscript, and gave final approval of the version to be published.

RC: assisted with the collection and analysis of data, made critical revisions to the manuscript, and gave final approval of the version to be published.

MG: assisted with the collection and analysis of data, made critical revisions to the manuscript, and gave final approval of the version to be published.

WG: assisted with the collection and analysis of data, made critical revisions to the manuscript, and gave final approval of the version to be published.

SHinze: assisted with the collection and analysis of data, made critical revisions to the manuscript, and gave final approval of the version to be published.

CN: assisted with the collection and analysis of data, made critical revisions to the manuscript, and gave final approval of the version to be published.

NRG: assisted with the collection and analysis of data, made critical revisions to the manuscript, and gave final approval of the version to be published.

AS: assisted with the collection and analysis of data, made critical revisions to the manuscript, and gave final approval of the version to be published.

GS: assisted with the collection and analysis of data, made critical revisions to the manuscript, and gave final approval of the version to be published.

SS: assisted with the collection and analysis of data, made critical revisions to the manuscript, and gave final approval of the version to be published.

TVL: assisted with the collection and analysis of data, made critical revisions to the manuscript, and gave final approval of the version to be published.

VW: assisted with the collection and analysis of data, made critical revisions to the manuscript, and gave final approval of the version to be published.

SW: assisted with the collection and analysis of data, made critical revisions to the manuscript, and gave final approval of the version to be published.

EY: assisted with the collection and analysis of data, made critical revisions to the manuscript, and gave final approval of the version to be published.

AZ: assisted with the collection and analysis of data, made critical revisions to the manuscript, and gave final approval of the version to be published.

DM: designed and conceptualized the study, assisted with the collection and analysis of data, made critical revisions to the manuscript, and gave final approval of the version to be published.

LW: provided input for the design and methodology, assisted with the collection and analysis of data, made critical revisions to the manuscript, and gave final approval of the version to be published.

SHaustein: designed and conceptualized the study, assisted with the collection and analysis of data, made critical revisions to the manuscript, and gave final approval of the version to be published.

## Acknowledgements

We gratefully acknowledge Anton Ninkov, Chantal Ripp, and Marc-Andre Simard for pilot-testing the Welphi survey for Round 1 of the Delphi process.

We gratefully acknowledge those who participated in Round 1 of the Delphi process, including: Per Ahlgren (Uppsala University, Sweden), Isola Ajiferuke (University of Western Ontario, Canada), Alberto Baccini (Università degli Studi di Siena, Italy), Caitlin Bakker (University of Regina, Canada), Aparna Basu (NISTADS, New Delhi (ex), India), Rachel Borchardt (American University, United States), Frédérique Bordignon (Ecole nationale des ponts et chaussées and LISIS, France), Bernardo Pereira Cabral (Federal University of Bahia, Brazil), David Campbell (Elsevier, Canada), Heather L. Coates (Indiana University, United States), Kyle Demes (OurResearch, Canada), Fereshteh Didegah (Karolinska Institutet, Sweden), Tim Engels (University of Antwerp, Belgium), Rabishankar Giri (Presidency University, Kolkata, India), Andrew Gray (University College London, United Kingdom), Sven E. Hug (Swiss Science Council, Switzerland), Brian D. Husbands (The Ohio State University, United States), Hamid R. Jamali (Charles Sturt University, Australia), Kayvan Kousha (University of Wolverhampton, United Kingdom), Geoff Krause (Dalhousie University, Canada), J. P. S. Kumaravel (Madurai Kamaraj University, India), Hiran H. Lathabai (Amrita Vishwa Vidyapeetham, India), Kun Lu (University of Oklahoma, United States), Fanrong Meng (Wageningen University & Research, Netherlands), Lokman I. Meho (American University of Beirut, Lebanon), Rogério Mugnaini (University of São Paulo, Brazil), José Luis Ortega (Spanish National Research Council, Spain), Joana Osório (Oxford PharmaGenesis, United Kingdom), Sheeba Pakkan (JSS Academy of Higher Education and Research, India), Saroj Kumar Panda (Central University of Rajasthan, India), Han Woo Park (YeungNam University, Republic of Korea), Cihan Papan (University Hospital Bonn, Germany), Ismael Rafols (CSIC & Leiden University, Spain), Amy Riegelman (University of Minnesota, United States), Imane Rhassate (School of Information Sciences, Morocco), Roberto Cruz Romero (German Centre for Higher Education Research and Science Studies (DZHW), Germany), Ronald Rousseau (KU Leuven and University of Antwerp, Belgium), Marion Schmidt (German Centre for Higher Education Research and Science Studies (DZHW), Germany), Eric Schares (Iowa State University, United States), João de Deus Barreto Segundo (Escola Bahiana de Medicina e Saúde Pública, Brazil), Francesca Soldati (University of Aberdeen, United Kingdom), Mike Thelwall (University of Sheffield, United Kingdom), Alyson Vaaler (Texas A&M University, United States), Dietmar Wolfram (University of Wisconsin-Milwaukee, United States), Houqiang Yu (Sun Yat-sen University, China), Alexey Zheleznov (European University at St. Petersburg, Russia), and Lin Zhang (Wuhan University, China).

We gratefully acknowledge Chantal Ripp for taking notes during Round 2 of the Delphi process.

We gratefully acknowledge Stephan Gauch at Humboldt University Berlin for providing space/location and technical equipment, and organizing the recording and transcription of Round 2 of the Delphi process.

## Figure Legend

**Figure 1:** Summary of the Methods Used to Develop the GLOBAL

**Figure 2:** Summary of the Results of the GLOBAL Development Process

## Table Legend

**Table 1**: Examples of Research Activity Units Used in Bibliometric Analyses

**Table 2**: Summary of Delphi Voting Rounds

**Table 3**: Round 1 and 2 Participant Demographic Characteristics

**Table 4**: Round 1 Delphi Online Consensus Group Themes

**Table 5:** Summary of Revisions to GLOBAL Items Post-Delphi Process

**Table 6:** GLOBAL Pilot-Test Themes

**Table 7**: The Final 28-item GLOBAL for the Reporting of Bibliometric Analyses

# Figures

## Figure 1: Summary of the Methods Used to Develop the GLOBAL

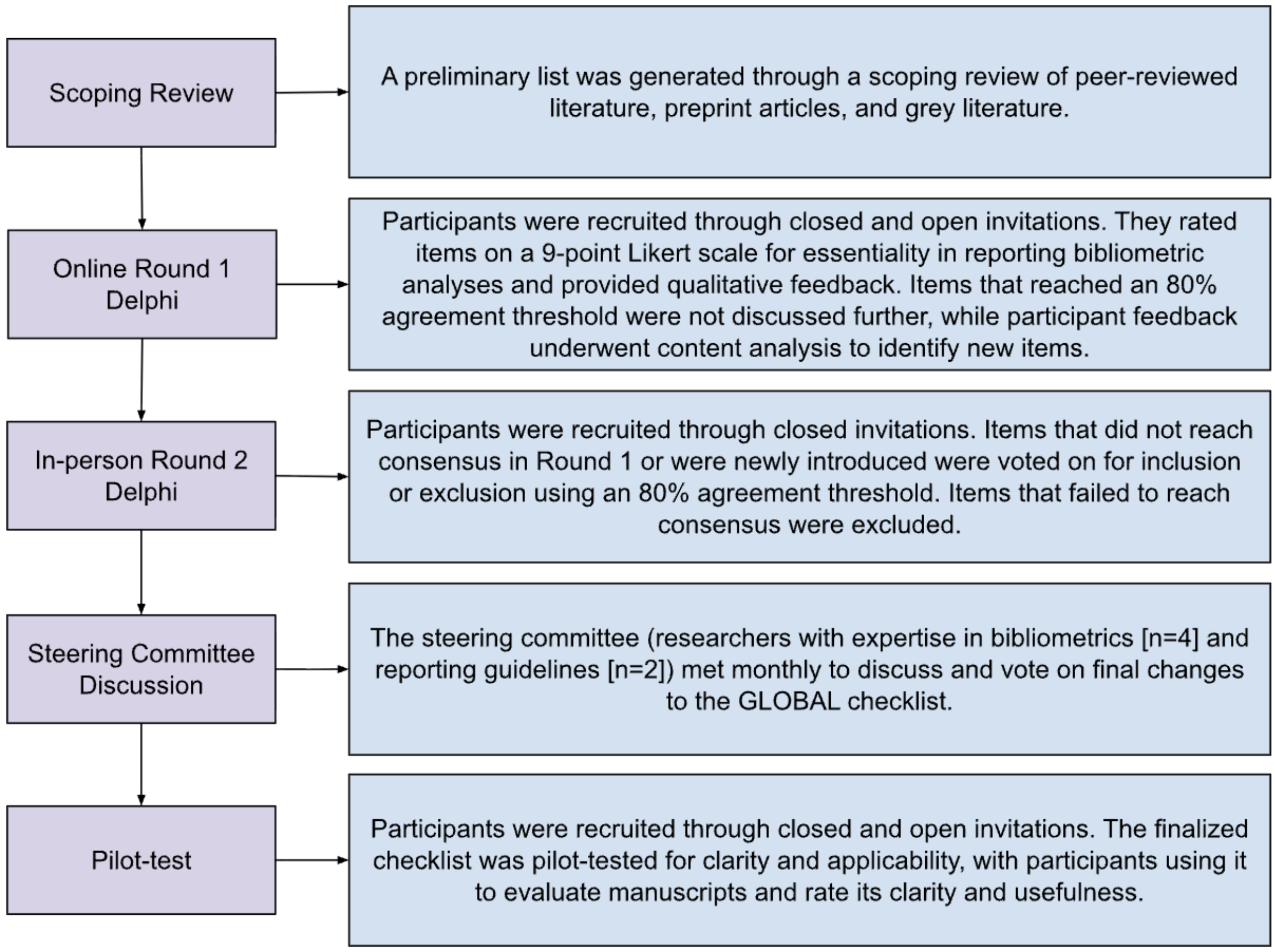

# Figure 2: Summary of the Results of the GLOBAL Development Process

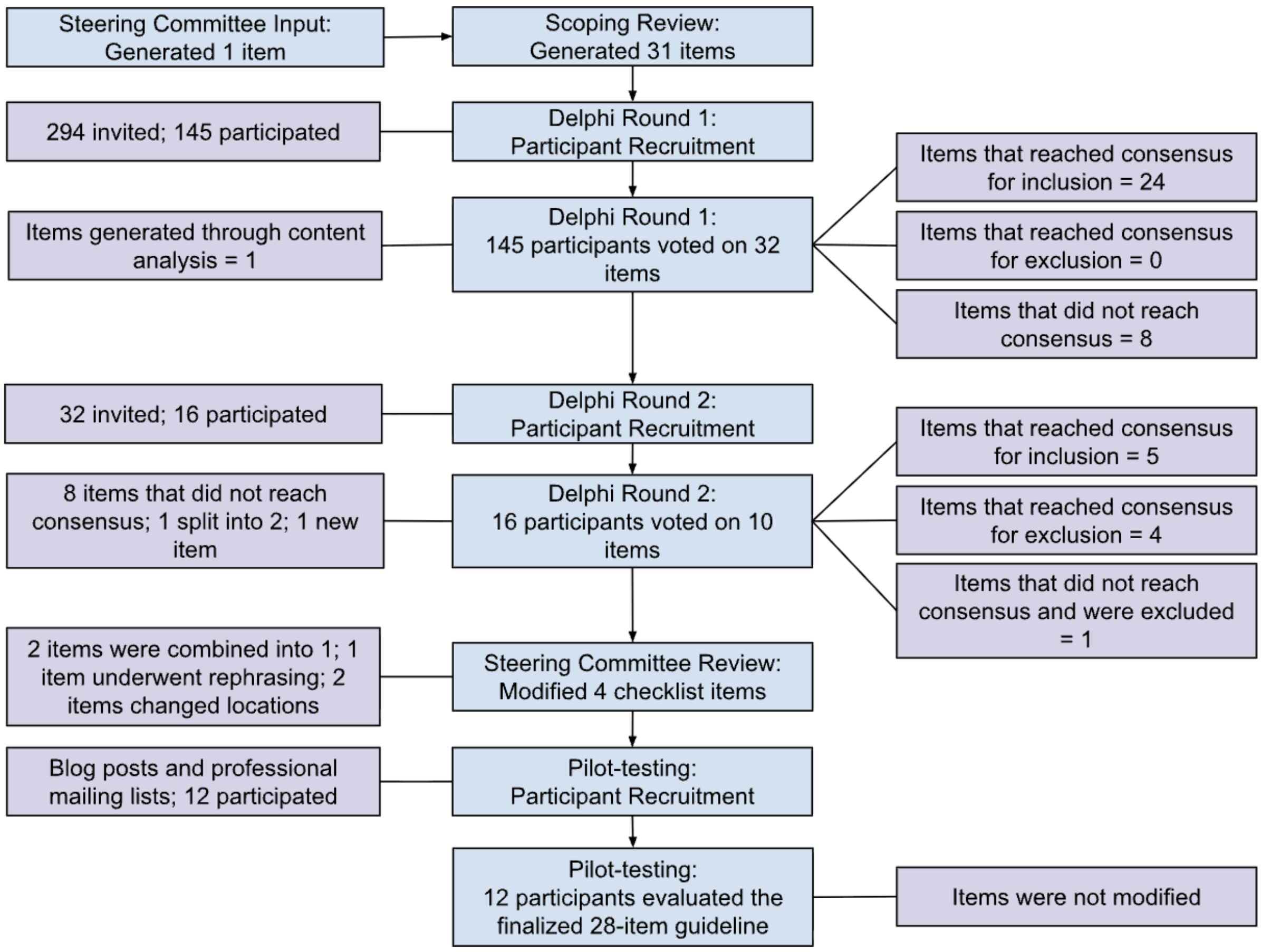

# Tables

## Table 1: Examples of Research Activity Units Used in Bibliometric Analyses

| | **Micro** | **Meso** | **Macro** |
|---|---|---|---|
| **People** | Researchers | Groups/Institutions | Countries/International Organizations |
| For example: | *Albert Einstein* | *Department of Mathematics/Princeton University* | *Switzerland/International Society for Informetrics and Scientometrics (ISSI)* |
| **Scope** | Topics | Areas/Fields | Disciplines |
| For example: | *Vaccine* | *Public Health* | *Medicine* |
| **Operation** | Projects | Programs | Portfolios |
| For example: | *The Human Genome Project* | *National Institute on Aging (NIA)* | *National Institutes of Health (NIH), National Science Foundation (NSF)* |
| **Output** | Single publications | Journals/Conferences | Global |
| For example: | *BERT 10.18653/v1/N19-1423* | Journal of the American Medical Association (JAMA) | Growth of science |
| **Impact** | Author citation impact | Journal citation impact / Institutional citation impact | Policy |
| For example: | Web of Science Author Impact Beamplot | Journal *impact factor / Share of papers belonging to the 10% most-frequently cited* | *Foresights or Delphi* |

## Table 2: Summary of Delphi Voting Rounds

| Section | Guideline item [a] | | | Agreement [b] (n [%]) | | Final Outcome |
|---|---|---|---|---|---|---|
| | Preliminary | Round 1 | Round 2 | Round 1 [c] | Round 2 [c] | |
| **Title** | In the title, identify the study as a bibliometric analysis and indicate the time period and key issues/topic. | In the title, identify the study as a bibliometric analysis and indicate the time period and key issues/topic. | In the title, identify the study as a bibliometric analysis and indicate the time period and key issues/topic. | Essential (1-3): 74 (51.03%) Neutral (4-6): 57 (39.31%) Non-Essential (7-9): 14 (9.66%) | Include: 3 (18.75%) **Exclude: 13 (81.25%)** Abstain: 0 (0%) | Excluded |
| **Abstract** | Abstract should be reflective of the bibliometric analysis, including scope, data collection, analysis, and results. | [Consensus Reached in Round 1] | [Consensus Reached in Round 1] | **Essential (1-3): 120 (83.33%)** Neutral (4-6): 23 (15.97%) Non-Essential (7-9): 1 (0.69%) | [Consensus Reached in Round 1] | Included |
| **Introduction** | Situate the bibliometric analysis within the context of relevant pre-existing literature, identifying the gap in literature. | [Consensus Reached in Round 1] | [Consensus Reached in Round 1] | **Essential (1-3): 117 (81.25%)** Neutral (4-6): 22 (15.28%) Non-Essential (7-9): 5 (3.47%) | [Consensus Reached in Round 1] | Included |
| **Introduction** | Define the aim, scope, rationale, and/or objective of the bibliometric analysis. | [Consensus Reached in Round 1] | [Consensus Reached in Round 1] | **Essential (1-3): 137 (95.80%)**<br>Neutral (4-6): 6 (4.20%)<br>Non-Essential (7-9): 0 (0.00%) | [Consensus Reached in Round 1] | Included |
| **Introduction** | Define the research question. | [Consensus Reached in Round 1] | [Consensus Reached in Round 1] | **Essential (1-3): 131 (91.61%)**<br>Neutral (4-6): 12 (8.39%)<br>Non-Essential (7-9): 0 (0.00%) | [Consensus Reached in Round 1] | Included |
| **Introduction** | Clearly define all relevant terms and definitions used within the bibliometric analysis. | Clearly define all relevant terms and definitions used within the bibliometric analysis. | <u>Explicitly specify</u> relevant terms, <u>concepts</u>, and <u>theoretical frameworks</u> used in the <u>study</u>. | Essential (1-3): 104 (72.73%)<br>Neutral (4-6): 35 (24.48%)<br>Non-Essential (7-9): 4 (2.80%) | **Include: 16 (100%)**<br>Exclude: 13 (81.25%)<br>Abstain: 0 (0%) | Included |
| **Introduction** | Describe the intended target audience of the bibliometric analysis (e.g., researchers, public, media, etc.). Describe the ways in which the information | Describe the intended target audience of the bibliometric analysis (e.g., researchers, public, media, etc.). Describe the ways in which the information | <u>[Rephrased into two items]</u><br><br>Item #1<br>Describe the intended target audience of the bibliometric analysis (e.g. researchers, | Essential (1-3): 56 (39.16%)<br>Neutral (4-6): 76 (53.15%)<br>Non-Essential (7-9): 11 (7.63%) | Include: 1 (6.25%)<br>**Exclude: 15 (93.75%)**<br>Abstain: 0 (0%) | Excluded |
| | included in the report may be used for the target audience. | included in the report may be used for the target audience. | public, media, etc). | | | |
| | | | [Rephrased into two items]<br><br>Item #2<br>Describe the ways in which the information included in the report is expected to be of relevance or intended to be used. | | Include:<br>3 (18.75%)<br>Exclude:<br>10 (62.5%)<br>Abstain:<br>3 (18.75%) | Excluded [d] |
| **Methods** | Describe the bibliometric methods used. | [Consensus Reached in Round 1] | [Consensus Reached in Round 1] | **Essential (1-3): 137 (95.74%)**<br>Neutral (4-6):<br>6 (4.26%)<br>Non-Essential (7-9):<br>0 (0.00%) | [Consensus Reached in Round 1] | Included |
| **Methods** | Define the units of analysis that are analysed (i.e., micro-, meso-, and macro-level) in the bibliometric analysis (e.g., countries, | [Consensus Reached in Round 1] | [Consensus Reached in Round 1] | **Essential (1-3): 126 (90.00%)**<br>Neutral (4-6):<br>14 (10.00%)<br>Non-Essential (7-9):<br>0 (0.00%) | [Consensus Reached in Round 1] | Included |
| | institutions, authors). | | | | | |
| **Methods** | Describe the bibliometric data collection methods, including any limitations. | [Consensus Reached in Round 1] | [Consensus Reached in Round 1] | **Essential (1-3): 137 (97.86%)**<br>Neutral (4-6): 3 (2.14%)<br>Non-Essential (7-9): 0 (0.00%) | [Consensus Reached in Round 1] | Included |
| **Methods** | Describe the databases and data sources used, including any limitations. | [Consensus Reached in Round 1] | [Consensus Reached in Round 1] | **Essential (1-3): 137 (97.86%)**<br>Neutral (4-6): 2 (1.43%)<br>Non-Essential (7-9): 1 (0.71%) | [Consensus Reached in Round 1] | Included |
| **Methods** | Present the full search strategies for all databases used, including any filters and limits that were applied. | [Consensus Reached in Round 1] | [Consensus Reached in Round 1] | **Essential (1-3): 124 (88.57%)**<br>Neutral (4-6): 16 (11.43%)<br>Non-Essential (7-9): 0 (0.00%) | [Consensus Reached in Round 1] | Included |
| **Methods** | Describe the data collection time frame. | [Consensus Reached in Round 1] | [Consensus Reached in Round 1] | **Essential (1-3): 126 (90.00%)**<br>Neutral (4-6): 14 (10.00%) | [Consensus Reached in Round 1] | Included |
| | | | | Non-Essential (7-9):<br>0 (0.00%) | | |
| **Methods** | Describe the search results and selection processes (e.g., inclusion/exclusion). If applicable, use a flow diagram. | [Consensus Reached in Round 1] | [Consensus Reached in Round 1] | **Essential (1-3): 114 (81.43%)**<br>Neutral (4-6):<br>26 (18.57%)<br>Non-Essential (7-9):<br>0 (0.00%) | [Consensus Reached in Round 1] | Included |
| **Methods** | Describe the data cleaning methods, including any limitations. [e] | [Consensus Reached in Round 1] | [Consensus Reached in Round 1] | **Essential (1-3): 120 (85.71%)**<br>Neutral (4-6):<br>19 (13.57%)<br>Non-Essential (7-9):<br>1 (0.71%) | [Consensus Reached in Round 1] | Included |
| **Methods** | Describe the bibliometric data analysis methods used. | [Consensus Reached in Round 1] | [Consensus Reached in Round 1] | **Essential (1-3): 134 (95.71%)**<br>Neutral (4-6):<br>6 (4.29%)<br>Non-Essential (7-9):<br>0 (0.00%) | [Consensus Reached in Round 1] | Included |
| **Methods** | Specify the analytical software used and the parameter settings selected. | [Consensus Reached in Round 1] | [Consensus Reached in Round 1] | **Essential (1-3): 116 (82.86%)**<br>Neutral (4-6):<br>24 (17.14%)<br>Non-Essential (7-9): | [Consensus Reached in Round 1] | Included |
| | | | | 0 (0.00%) | | |
| **Methods** | Describe the bibliometric indicators used. | [Consensus Reached in Round 1] | [Consensus Reached in Round 1] | **Essential (1-3): 128 (91.43%)** Neutral (4-6): 12 (8.57%) Non-Essential (7-9): 0 (0.00%) | [Consensus Reached in Round 1] | Included |
| **Methods** | If applicable, define the calculations/formulas used for indicators in the bibliometric analysis. | [Consensus Reached in Round 1] | [Consensus Reached in Round 1] | **Essential (1-3): 115 (82.14%)** Neutral (4-6): 25 (17.86%) Non-Essential (7-9): 0 (0.00%) | [Consensus Reached in Round 1] | Included |
| **Methods** | Provide sufficient detail in the bibliometric analysis manuscript to ensure full replicability/transparency of methods. | [Consensus Reached in Round 1] | [Consensus Reached in Round 1] | **Essential (1-3): 118 (84.29%)** Neutral (4-6): 22 (15.71%) Non-Essential (7-9): 0 (0.00%) | [Consensus Reached in Round 1] | Included |
| **Results** | Describe the results and key findings. | [Consensus Reached in Round 1] | [Consensus Reached in Round 1] | **Essential (1-3): 136 (97.14%)** Neutral (4-6): 4 (2.86%) Non-Essential (7-9): | [Consensus Reached in Round 1] | Included |
| | | | | 0 (0.00%) | | |
| **Results** | Describe the results of bibliometric analysis techniques used. | [Consensus Reached in Round 1] | [Consensus Reached in Round 1] | **Essential (1-3): 123 (87.86%)**<br>Neutral (4-6): 15 (10.71%)<br>Non-Essential (7-9): 2 (1.43%) | [Consensus Reached in Round 1] | Included |
| **Results** | Visualize the results through the use of figures, graphs, and/or tables. Ensure the visualizations are simple and easy to interpret. Aesthetic bibliometric visualization should not replace a rigorous bibliometric analysis. | Visualize the results through the use of figures, graphs, and/or tables. Ensure the visualizations are simple and easy to interpret. Aesthetic bibliometric visualization should not replace a rigorous bibliometric analysis. | Ensure figures, tables and visualizations clarify and/or facilitate the interpretation of the results without misleading. | Essential (1-3): 102 (72.86%)<br>Neutral (4-6): 33 (23.57%)<br>Non-Essential (7-9): 5 (3.57%) | **Include: 13 (86.67%)**<br>Exclude: 0 (0%)<br>Abstain: 2 (13.33%) | Included |
| **Results** | If applicable, report the uncertainty/dispersion/heterogeneity depending on the type of analysis and error values of | If applicable, report the uncertainty/dispersion/heterogeneity depending on the type of analysis and error values of | If appropriate, report the uncertainty/dispersion/heterogeneity depending on the type of data and analysis, and error values of | Essential (1-3): 97 (69.29%)<br>Neutral (4-6): 43 (30.71%)<br>Non-Essential (7-9): 0 (0.00%) | **Include: 12 (80%)**<br>Exclude: 2 (13.33%)<br>Abstain: 1 (6.67%) | Included |
| | bibliometric indicators. | bibliometric indicators. | bibliometric indicators | | | |
| **Discussion** | Summarize and discuss study findings. | [Consensus Reached in Round 1] | [Consensus Reached in Round 1] | **Essential (1-3): 129 (92.14%)** Neutral (4-6): 11 (7.86%) Non-Essential (7-9): 0 (0.00%) | [Consensus Reached in Round 1] | Included |
| **Discussion** | Elaborate on the applicability and implications of study findings. | Elaborate on the applicability and implications of study findings. | Discuss the applicability and implications of study findings. | Essential (1-3): 105 (75.00%) Neutral (4-6): 34 (24.29%) Non-Essential (7-9): 1 (0.71%) | Include: 9 (60%) Exclude: 5 (33.33%) Abstain: 1 (6.67%) | Excluded [d] |
| **Discussion** | Provide context for the results of the bibliometric analysis and situate the study findings in existing literature. | Provide context for the results of the bibliometric analysis and situate the study findings in the existing literature. | Provide context for and situate the study findings in the literature. | Essential (1-3): 104 (74.29%) Neutral (4-6): 35 (25.00%) Non-Essential (7-9): 1 (0.71%) | **Include: 13 (86.67%)** Exclude: 2 (13.33%) Abstain: 0 (0%) | Included |
| **Discussion** | Discuss the strengths, limitations, and potential biases of | [Consensus Reached in Round 1] | [Consensus Reached in Round 1] | **Essential (1-3): 128 (90.00%)** Neutral (4-6): 13 (9.29%) | [Consensus Reached in Round 1] | Included |
| | the bibliometric analysis. | | | Non-Essential (7-9): 1 (0.71%) | | |
| **Discussion** | Identify future directions for research. | Identify future directions for research | Identify future directions for research | Essential (1-3): 54 (38.57%) Neutral (4-6): 79 (56.43%) Non-Essential (7-9): 7 (5.00%) | Include: 3 (20%) **Exclude: 12 (80%)** Abstain: 0 (0%) | Excluded |
| **Other** | Disclose any existing or potential conflicts of interest and/or sources of financial or non-financial support. | [Consensus Reached in Round 1] | [Consensus Reached in Round 1] | **Essential (1-3): 112 (80.00%)** Neutral (4-6): 27 (19.29%) Non-Essential (7-9): 1 (0.71%) | [Consensus Reached in Round 1] | Included |
| **Other** | Describe the availability and accessibility of data. | [Consensus Reached in Round 1] | [Consensus Reached in Round 1] | **Essential (1-3): 114 (81.43%)** Neutral (4-6): 26 (18.57%) Non-Essential (7-9): 0 (0.00%) | [Consensus Reached in Round 1] | Included |
| **Other** | Use references and citations to support statements and methods used. | [Consensus Reached in Round 1] | [Consensus Reached in Round 1] | **Essential (1-3): 125 (89.29%)** Neutral (4-6): 15 (10.71%) Non-Essential (7-9): | [Consensus Reached in Round 1] | Included |
| | | | | 0 (0.00%) | | |
| **Other** | [Not Included in Round 1] | Provide a clear study materials and data sharing statements (e.g. if datasets, data sources, codes used for the analysis, software, and/or calculations are provided or not). | Provide a statement about whether study materials, data and/or code are shared and if so, where and how it can be accessed. | [Not Included in Round 1] | **Include: 14 (100%)** Exclude: 0 (0%) Abstain: 0 (0%) | Included |

[a] Underlining denotes text changes made between rounds.
[b] **Bold** indicates consensus. Consensus is defined as 80% of participants agree.
[c] Round 1 items were scored on a 9-point Likert scale, where 1 to 3 points were categorized as 'essential', 4 to 7 points were categorized as 'neutral,' and 7 to 9 points were categorized as 'non-essential' for inclusion within the tool. Round 2 items were scored using 'include in checklist', 'exclude from checklist', and 'abstain from voting' for inclusion within the tool.
[d] Item excluded because the 80% threshold for consensus was not reached.
[e] Item included based on steering committee discussions, not the literature.

## Table 3: Round 1 and 2 Participant Demographic Characteristics

| Demographic | Participant characteristics | Responses (n (%)) | |
|---|---|---|---|
| | | Round 1 | Round 2 |
| Gender | Male | 91 (62.8%) | 11 (68.8%) |
| | Female | 47 (28.9%) | 4 (25.0%) |
| | Prefer not to say | 7 (4.8%) | 1 (6.2%) |
| | | N=145 | N=16 |
| | 25-29 | 2 (1.5%) | 0 (0.0%) |
| Age | 30-34 | 12 (8.9%) | 3 (18.8%) |
| | 35-39 | 28 (20.7%) | 2 (12.5%) |
| | 40-44 | 28 (20.7%) | 2 (12.5%) |
| | 45-49 | 16 (11.8%) | 1 (6.2%) |
| | 50-54 | 15 (11.1%) | 2 (12.5%) |
| | 55-59 | 13 (9.6%) | 3 (18.8%) |
| | 60-64 | 8 (5.9%) | 1 (6.2%) |
| | 65-69 | 6 (4.4%) | 2 (12.5%) |
| | 70-74 | 4 (3.0%) | 0 (0.0%) |
| | 80+ | 2 (1.5%) | 0 (0.0%) |
| | Prefer not to say | 1 (0.7%) | 0 (0.0%) |
| | | N=135 | N=16 |
| Race/Ethnicity | East Asian | 17 (12.7%) | 1 (7.1%) |
| | Latino | 9 (6.7%) | 0 (0.0%) |
| | Middle Eastern | 5 (3.7%) | 0 (0.0%) |
| | South Asian | 6 (4.8%) | 0 (0.0%) |
| | White | 80 (59.7%) | 13 (92.9%) |
| | Black | 4 (3.0%) | 0 (0.0%) |
| | Southeast Asian | 3 (2.2%) | 0 (0.0%) |
| | Other | 3 (2.2%) | 0 (0.0%) |
| | Prefer not to say | 7 (5.2%) | 0 (0.0%) |
| | | N=134 | N=14 |
| Country of Work | Canada | 15 (12.6%) | 0 (0.0%) |
| | Germany | 9 (7.6%) | 4 (28.6%) |
| | Netherlands | 8 (6.7%) | 2 (14.3%) |
| | United Kingdom | 9 (7.6%) | 0 (0.0%) |
| | United States | 19 (16.0%) | 1 (7.1%) |
| | Other | 56 (47.1%) | 7 (50.0%) |
| | Prefer not to say | 3 (2.5%) | 0 (0.0%) |
| | | N=119 | N=14 |
| Role | Bibliometrician | 48 (22.6%) | 7 (26.9%) |
| | Librarian/Information Specialist | 36 (17.0%) | 3 (11.5%) |
| | Research Coordinator | 13 (6.1%) | 2 (7.7%) |
| | Associate Professor | 18 (8.5%) | 3 (11.5%) |
| | Full professor | 17 (8.0%) | 2 (7.7%) |
| | Other | 80 (37.7%) | 9 (32.6%) |
| | | N=212[a] | N=26[a] |
| Career Stage | Early career (no more than 5 years) | 17 (14.1%) | 1 (7.1%%) |
| | Mid-career (between 5 and 10 years) | 31 (25.6%) | 3 (21.4%) |
| | Senior career (more than 10 years) | 70 (57.9%) | 10 (71.4%) |
| | Prefer not to say | 3 (2.5%) | 0 (0.0%) |
| | | N=121 | N=14 |

[a] Participants chose more than one option.

## Table 4: Round 1 Delphi Online Consensus Group Themes

| Theme | Codes | Quotes |
|---|---|---|
| GLOBAL Objectives | Clarify Purpose | "Why is this needed at this time?" (P15)<br>"The meaning of 'reporting bibliometric studies' could be broad. It is unclear whether it specifically apply to reports, [...] research articles, or any document based on a bibliometric analysis." (P77) |
| | Adapt to Audience | "I wonder if the reporting needs to be adapted to the audience" (P156)<br>"I sometimes found it difficult to answer the questions because I produce different types of analysis for very different audiences" (P122) |
| | Extensions | "How about other types of reporting, e.g. benchmarking reports that institutions and governments use" (P342) |
| GLOBAL Formatting and Structure | Redundant/Generic Requirements | "Some of the items seem to be overlapping in meaning, causing unnecessary redundancy" (P241)<br>"Many of these items seem not particular to bibliometric studies but rather standard elements of journal articles" (P378) |
| | Include Examples | "It would be nice to see some "recipes" (representative examples)." (P320) |
| | Item Editing/Suggestions | "[...] the question of data availability and the conflict of interest [...] should be better defined." (P25)<br>"By making all this mandatory one runs the risk of making papers heavy and impenetrable" (P360) |

## Table 5: Summary of Revisions to GLOBAL Items Post-Delphi Process

| Change Type | Original Item(s) | Revised Items | Rationale |
|---|---|---|---|
| Item Reorder | Provide any existing or potential conflicts of interest and/or sources of financial or non-financial support. | Use references and citations to support statements and methods used. | The items were reordered to improve reporting flow and follow the order commonly found in research papers. |
| | Describe the availability and accessibility of data. | Describe the availability and accessibility of data. | |
| | Use references and citations to support statements and methods used. | Provide a statement on whether study materials, data, and/or code are shared, and if so, where and how they can be accessed. | |
| | Provide a statement on whether study materials, data, and/or code are shared, and if so, where and how they can be accessed. | Provide any existing or potential conflicts of interest and/or sources of financial or non-financial support. | |
| Item Merge | Describe the availability and accessibility of data. | Describe the availability and accessibility of data by providing a statement on whether study materials, data, and/or code are shared, and if so, where and how they can be accessed. | The two items were merged into one to remove overlapping expectations related to data sharing and accessibility. |
| | Provide a statement on whether study materials, data, and/or code are shared, and if so, where and how they can be accessed. | | |
| Word Change | Use references and citations to support statements and methods used. | Use references to support statements and methods used. | ”Citations” was removed to avoid confusion between outgoing and incoming citations commonly used in bibliometrics. "References" more clearly refers to sources cited in the current study. |

## Table 6: GLOBAL Pilot-Test Themes

| Theme | Codes | Quotes |
|---|---|---|
| GLOBAL Clarity and Applicability | Edit (include/remove) wording | "Encourage authors to distinguish between size-dependent and size-independent indicators." (Item 3.1) (P5) |
| | Current wording is limiting/restrictive | "I believe the study goal can be expressed differently, rather than as an explicit set of research questions. The paper I am reviewing demonstrates this effectively." (Item 2.3) (P6) |
| | Current wording lacks specificity/clarity | "Should authors provide formulas in the text or in an appendix? What source should be cited to provide standard definitions? A scientific article or a service provider FAQ? (e.g., the FWCI formula)." (Item 3.12) (P2) |
| | Include definitions | "what is a bibliiometric indicators and what is not?" (Item 3.11) (P1) |
| | Include more examples | "The example in E&E text in this item falls under 'Humanities and Social sciences'. Journals dealing with STEM might not entertain works with less methodical and analytical rigor. It is suggested to include examples from multiple fields." (Item 3.1) (P9) |
| GLOBAL Structure and Format | Overlap between items | "I think there is an overlap with 3.3. Can the two be merged? Or can it be clarified that one is about data collection and the other about data sources?" (Item 3.4) (P12) |
| | Change location of item | "I preferred specifying relevant terms and concepts in the method sections together with more technical definitions." (Item 2.4) (P7) |
| | Not specific to bibliometrics | "it applies to any type of research, whether bibliometrics or not" (Item 2.1) (P1) |

## Table 7: The Final 28-item GLOBAL for the Reporting of Bibliometric Analyses

| Reporting Items | |
|---|---|
| **Abstract** | |
| 1.1 | Abstract should be reflective of the bibliometric analysis, including scope, data collection, analysis, and results. |
| **Introduction** | |
| 2.1 | Situate the bibliometric analysis within the context of relevant pre-existing literature, identifying the gap in literature. |
| 2.2 | Define the aim, scope, rationale, and/or objective of the bibliometric analysis. |
| 2.3 | Define the research question. |
| 2.4 | Explicitly specify relevant terms, concepts, and theoretical frameworks used in the study. |
| **Methods** | |
| 3.1 | Describe the bibliometric methods used. |
| 3.2 | Define the units of analysis that are analysed (i.e., micro-, meso-, and macro-level) in the bibliometric analysis (e.g., countries, institutions, authors). |
| 3.3 | Describe the bibliometric data collection methods, including any limitations. |
| 3.4 | Describe the databases and data sources used, including any limitations. |
| 3.5 | Present the full search strategies for all databases used, including any filters and limits that were applied. |
| 3.6 | Describe the data collection time frame. |
| 3.7 | Describe the search results and selection processes (e.g., inclusion/exclusion). If applicable, use a flow diagram. |
| 3.8 | Describe the data cleaning methods, including any limitations. |
| 3.9 | Describe the bibliometric data analysis methods used. |
| 3.10 | Specify the analytical software used and the parameter settings selected. |
| 3.11 | Describe the bibliometric indicators used. |
| 3.12 | If applicable, define the calculations/formulas used for indicators in the bibliometric analysis. |
| 3.13 | Provide sufficient detail in the bibliometric analysis manuscript to ensure full replicability/transparency of methods. |
| **Results** | |
| 4.1 | Describe the results and key findings. |
| 4.2 | Describe the results of bibliometric analysis techniques used. |
| 4.3 | Ensure figures, tables and visualizations clarify and/or facilitate the interpretation of the results without misleading. |
| 4.4 | If appropriate, report the uncertainty/dispersion/heterogeneity depending on the type of data and analysis, and error values of bibliometric indicators. |
| **Discussion** | |
| 5.1 | Summarize and discuss study findings. |
| 5.2 | Provide context for and situate the study findings in the literature. |
| 5.3 | Discuss the strengths, limitations, and potential biases of the bibliometric analysis. |
| **Other** | |
| 6.1 | Use references to support statements and methods used. |
| 6.2 | Describe the availability and accessibility of data by providing a statement about whether study materials, data and/or code are shared and if so, where and how it can be accessed. |
| 6.3 | Disclose any existing or potential conflicts of interest and/or sources of financial or non-financial support. |